# Fabric Softener/Cellulose Nanocrystal Interaction: A Model For Assessing Surfactant Deposition On Cotton

E.K. Oikonomou*[,1], F. Mousseau[1], N. Christov[2], G. Cristobal[2], A. Vacher[3],
M. Airiau[3], C. Bourgaux[4], L. Heux[5], and J.-F. Berret*[,1]

[1]Laboratoire Matière et Systèmes Complexes, UMR 7057 CNRS Université Denis Diderot Paris-VII,
Bâtiment Condorcet, 10 rue Alice Domon et Léonie Duquet, 75205 Paris, France
[2]Solvay Research & Innovation Center Singapore, 1 Biopolis Drive, Amnios, Singapore 138622
[3]Solvay Research & Innovation Centre Paris, 52 rue de la Haie Coq, 93306 Aubervilliers cedex
[4]Institut Galien Paris-Sud - UMR CNRS 8612, Faculté de Pharmacie, Université Paris-Sud XI, 92296
Châtenay-Malabry Cedex, France
[5]Centre de recherches sur les macromolécules végétales, BP 53, 38041 Grenoble cedex 9, France

**Abstract:** There is currently a renewed interest for improving household and personal care formulations to provide more environment friendly products. Fabric conditioners used as softeners have to fulfill a number of stability and biodegradability requirements. They should also display significant adsorption on cotton in the conditions of use. The quantification of surfactant adsorption remains however difficult because the fabric woven structure is complex and deposited amounts are generally small. Here we propose a method to evaluate cellulose/surfactant interactions with increased detection sensitivity. The method is based on the use of cellulose nanocrystals in *lieu* of micron-sized fibers or yarns, combined with different techniques including light scattering, optical and electron microscopy, and electrophoretic mobility. Cellulose nanocrystals are rod-shaped nanoparticles in the form of 200 nm laths that are negatively charged and can be dispersed in bulk solutions. In this work, we use a double-tailed cationic surfactant present in fabric softener. Results show that the surfactants self-assemble into unilamellar, multivesicular and multilamellar vesicles, and interaction with cellulose nanocrystals is driven by electrostatics. Mutual interactions are strong and lead to the formation of large-scale aggregates, where the vesicles remain intact at the cellulose surface. The technique developed here could be exploited to rapidly assess fabric conditioner efficiency obtained by varying the nature and content of their chemical additives.

*Corresponding authors: evdokia.oikonomou@univ-paris-diderot.fr*

*jean-francois.berret@univ-paris-diderot.fr*

## I - Introduction

Surfactant adsorption at the solid-liquid interface has been a subject of considerable attention in the last decades. Interfaces usually modify surfactant aggregation mechanisms and surfactants bring novel properties to surfaces in modulating their interaction, solubility or surface energy. The main applications of surfactant at interfaces are found in home and personal care formulations, paper manufacturing and recycling, soil remediation and water treatment. For single alkyl chains, deposition takes place via mechanisms that are now well understood. Surfactants adsorb under the form of single monomer units, hemimicelles or uniform bilayers (admicelles)[1-4] and adsorption isotherms can be predicted.[3,5] In contrast, double-tailed surfactant forming vesicular systems have received comparatively less attention.[6] The deposition scenarios of surfactant vesicles on planar substrates display some similarities with those found for phospholipids or membranes of biological origin.[7-8] These scenarios assume *i)* either the formation of a supported bilayer associated with a reorganization of the lipids towards the interface or *ii)* the formation of a supported vesicular layer observed when adsorbed vesicles maintain their structure. Supported lipid bilayers on plane surfaces have been extensively studied in the context of surface patterning and stability and more recently of resistance towards protein adsorption.[8]

Supported vesicular layers on solid surfaces have been seldom reported.[9-10]

For a variety of household and personal care products, the deposition of vesicles on solid substrates can have important consequences. Double-tailed cationic surfactants are for instance the major ingredients of fabric enhancers and hair conditioners. For fabrics, surfactants are used in the last rinse of the washing cycle during which they adsorb onto cellulose fibers and yarns. Fabric softener formulations fulfill a large number of requirements, which are related to their phase and temperature stability, non-toxicity, biodegrability and rheological properties.[11-16] One important requirement is that the surfactants self-assemble into nanometer- to micron-sized vesicles.[7,17-19] Nowadays, a renewed interest in conditioning formulations, either for hair or fabrics has emerged. This interest is triggered by the need to redefine eco-friendly products issued from natural or less aquatoxic origin.

Fabric softeners have been used and studied for decades, however the deposition process onto cotton fibers and the softening mechanism are not fully understood. Using streaming potential and fluorescence microscopy measurements, Kumar *et al.* investigated vesicle interaction with micrometer thick cotton fibers. These authors have found that the deposition process is electrostatically driven and that intact vesicles adsorb on cotton and viscose fibers.[7] Concerning the softening mechanism, it is generally admitted that adsorbed surfactants form a boundary layer after drying that lubricates the fibers and reduce friction, giving the fabric a soft feel to the touch.[11,14-15,20-21] Alternative mechanisms were also proposed. Igarashi *et al.* observed that hardness appears only following natural drying, but not if drying is performed under vacuum at a pressure around $10^3$ Pa.[12-13] For these authors, hardness originates from hydrogen bonding interactions that are mediated by bound water molecules, and that the softening effect comes from H-bonding reduction.[12-13]

The quantification of surfactant adsorption on cotton fibers and yarns is generally performed to investigate the nature and strength of interactions. The measurement of surfactant deposition remains however difficult because the fabric woven structure is complex and deposited amounts are generally small. In the present paper we propose a method to evaluate cellulose/surfactant interactions with increased detection sensitivity. This method combines the use of cellulose nanocrystals (CNC) in *lieu* of micron-sized fibers with that of light scattering. CNC are rod-shaped nanoparticles in the form of laths of length 200 nm and width 10 nm. The particles can be dispersed in water in various physicochemical conditions, making them ideal for bulk studies. CNC is a renewable, biodegradable, nontoxic nanomaterial obtained from natural cellulose.[6,22-23] It was shown recently that CNC strongly interact with single alkyl chain cationic surfactants via electrostatic interaction, leading to the formation of hemi and admicelles structures, as well as coacervate phases.[2]

In this work, the cellulose/surfactant interaction was monitored using the Continuous Variation Method developed for enzymatic activity evaluation in the late 1920's.[24] The technique was later adapted for the turbidity titration of colloids and polymers.[25-27] The Continuous Variation Method is based on light scattering measurements of mixed dispersions as a function of the mole fraction of the two components. Here we use the double-tailed surfactant ethanaminium, 2-hydroxyN,N-bis(2-hydroxy-ethyl)-N-methyl-esters (abbreviated as TEQ) which enters softener formulations already available on the market. Combination of light scattering, small and wide-angle x-ray scattering, optical and cryogenic electron microscopy allows us to show that TEQ self assemble into unilamellar, multivesicular and multilamellar vesicles, and that interaction with cellulose nanocrystal is driven by electrostatics.[28] It is also demonstrated that in the dilute concentration regime the deposition occurs via the adsorption of intact vesicles.

## II – Experimental Section

**Materials.** The esterquat surfactant ethanaminium, 2-hydroxyN,N-bis(2-hydroxye-thyl)-N-methyl-esters with saturated and unsaturated C16-18 aliphatic chains, abbreviated TEQ in the following was kindly provided by Solvay®. The counterions associated to the quaternized amines are methyl sulfate groups. The guar hydrocolloid dispersions were obtained from Solvay®. Nanocrystalline cellulose was prepared according to earlier reports using catalytic and selective oxidation.[29] Briefly, cotton linters provided by Buckeye Cellulose Corporation were hydrolyzed according to the method described by Revol *et al.*[30] treating the cellulosic substrate with 65% (w/v) sulfuric acid at 63 °C during 30 min. The suspensions were washed by repeated centrifugations, dialyzed against distilled water until neutrality and sonicated for 4 min with a Branson B-12 sonifier equipped with a 3 mm microtip. The suspensions were then filtered through 8 µm and then 1µm cellulose nitrate membranes (Whatman). At the end of the process, ~ 2 wt. % aqueous stock suspensions were obtained. The resulting nanoparticles are in the form of laths and have a length of around 180 ± 30 nm for their length and 17 ± 4 nm for their width for a height estimated around 7 nm.[22] For the fluorescent microscopy, PKH67 Fluorescent Cell (Aldrich) was diluted and mixed to the surfactant solutions. Water was deionized with a Millipore Milli-Q Water system. All the products were used without further purification.

**Light scattering.** The scattering intensity $I_S$ and hydrodynamic diameter $D_H$ were measured using the Zetasizer Nano ZS spectrometer (Malvern Instruments, Worcestershore, UK). A 4 mW He−Ne laser beam ($\lambda = 633$ nm) is used to illuminate the sample dispersion, and the scattered intensity is collected at a scattering angle of 173°. The Rayleigh ratio $\mathcal{R}$ was derived from the intensity according to the relationship: $\mathcal{R} = (I_S - I_w)n_0^2 \mathcal{R}_T / I_T n_T^2$ where $I_w$ and $I_T$ are the water and toluene scattering intensities respectively, $n_0 = 1.333$ and $n_T = 1.497$ the solution and toluene refractive indexes, and $\mathcal{R}_T$ the toluene Rayleigh ratio at $\lambda = 633$ nm ( $\mathcal{R}_T = 1.352 \times 10^{-5}\ cm^{-1}$ ) . The second-order autocorrelation function $g^{(2)}(t)$ is analyzed using the cumulant and CONTIN algorithms to determine the average diffusion coefficient $D_C$ of the scatterers. The hydrodynamic diameter is then calculated according to the Stokes-Einstein relation, $D_H = k_B T / 3\pi\eta D_C$ , where $k_B$ is the Boltzmann constant, $T$ the temperature and $\eta$ the solvent viscosity. Measurements were performed in triplicate at 25 °C after an equilibration time of 120 s.

**Zeta Potential**. Laser Doppler velocimetry (Zetasizer, Malvern Instruments, Worcestershore, UK) using the phase analysis light scattering mode and detection at an angle of 16° was performed to determine the electrophoretic mobility and zeta potential of the different dispersions studied. Measurements were performed in triplicate at 25 °C, after 120 s of thermal equilibration.

**X-ray scattering.** Samples were loaded into quartz capillaries (diameter 1.5 mm, Glass Müller, Berlin, Germany). The top of the capillaries was sealed with a drop of paraffin to prevent water evaporation. The temperature of the sample holder was set to 25°C. The scattered intensity was reported as a function of the scattering vector $q = 4\pi sin\theta/\lambda$, where $2\theta$ is the scattering angle and $\lambda$ the wavelength of the incident beam. Small-Angle X-ray Scattering (SAXS) experiments were performed on the SWING beamline at the SOLEIL synchrotron source (Saint Aubin, France). SAXS patterns were recorded using a two-dimensional AVIEX CCD detector placed in a vacuum detection tunnel. The beamline energy was set at 12 keV and the sample-to-detector distance was fixed to cover the $0.08 - 8$ nm$^{-1}$ $q$-range. The calibration was carried out with silver behenate. The acquisition time of each pattern was 250 ms; 5 acquisitions were averaged for each sample. Intensity values were normalized to account for beam intensity, acquisition time and sample transmission. Each scattering pattern was then integrated circularly to yield the intensity as a function of the wave-vector. The scattered intensity from a capillary filled with water was subtracted from the sample scattering curves. Wide-Angle X-ray Scattering (WAXS) experiments were performed on the Austrian beamline at the ELETTRA synchrotron source (Trieste, Italy) operated at 8 keV. The data were collected using a position sensitive linear gas detector filled with a mixture of argon and ethane. Exposure times were 300 s.

**Optical Microscopy.** Phase-contrast were acquired on an IX73 inverted microscope (Olympus) equipped with 20×, 40×, and 60× objectives. Seven microliters of TEQ dispersion were deposited on a glass plate and sealed into a Gene Frame (Abgene/ Advanced Biotech) dual adhesive system. An Exi-Blue camera (QImaging) and Metaview software (Universal Imaging Inc.) were used as the acquisition system.

**Fluorescent Microscopy**. For fluorescent microscopy, 100 µl of the TEQ surfactant solution were mixed with 100 µl of 1 µM PKH67 (Sigma) solution. PKH67 is used in cellular biology as a green fluorescent molecular linker developed for cell membrane labeling. It is characterized by a absorption maximum at 490 nm and an excitation maximum at 502 nm. The solution was then stirred and let equilibrate for 1 h before use.

**Cryo-TEM**. A few microliters of the samples were deposited on a lacey carbon coated 200 mesh (Ted Pella). The drop was blotted with a filter paper on a VitrobotTM (FEI) and the grid was quenched rapidly in liquid ethane, cooled with liquid nitrogen, to avoid the crystallization of the aqueous phase. The membrane was finally transferred into the vacuum column of a TEM microscope (JEOL 1400 operating at 120 kV) where it was maintained at liquid nitrogen temperature thanks to a cryo holder (Gatan). The magnification was selected between 3,000x and 40,000x, and images were recorded with an 2k-2k Ultrascan camera (Gatan).

**Mixing protocol.** The surfactant stock solutions were formulated at 60 °C while the pH was adjusted at 4.1 - 4.5. The interactions between cellulose nanocrystals and TEQ surfactants were investigated using the direct mixing formulation pathway.[31-33] Batches of CNC and TEQ surfactant were prepared in the same conditions of pH (pH 4.5) and concentration ($c$ = 0.01 and 0.1 wt. %) and then mixed at different ratios, noted $X = V_{Surf}/V_{CNC}$, where $V_{Surf}$ and $V_{CNC}$ denote the volumes of the TEQ surfactant and the CNC particles dispersions, respectively. After mixing at room temperature (25 °C), the dispersions were stirred rapidly, let to equilibrate for 5 minutes and the scattered intensity and hydrodynamic diameter were measured in triplicate. As the concentrations of the stock solutions are identical, the volumetric ratio $X$ is equivalent to the mass ratio between constituents. In the figures, the surfactant and cellulose nanocrystals stock solutions are set at $X = 10^{-3}$ and $X = 10^{3}$ respectively.

## III - Results and Discussion

### III.1 - Surfactant assembly at the nanometer scale

#### III.1.1 - Cryo-transmission electron microscopy

Aqueous TEQ dispersions were studied by cryo-TEM at concentrations relevant for the applications. Representative images obtained in the dilute regime (0.1%) and intermediate regimes (1 and 4 wt. %) are displayed in Figs. 1a, 1b and 1c respectively. Images show that the surfactants self-assemble locally into a bilayer structure. Fig. 1d represents the distribution of membrane thicknesses, resulting in a median value of 5.2 ± 0.6 nm (inset). This value is slightly higher than that of a regular C18 bilayer, and the difference could be ascribed to resolution issues specific to the Cryo-TEM technique.

At larger scale, the bilayers close on themselves to form unilamellar and multivesicular vesicles (indicated by arrows) of size between 100 nm and 1 µm.[34] Multivesicular vesicles are defined as a membrane compartment that encapsulates several smaller vesicles. At 4 wt. %, the vesicles are flattened by the confined geometry imposed by the cryo-TEM grid and as such appear non-spherical. Statistical analyses on vesicle samples show that unilamellar vesicles are mostly present at low concentration, whereas multivesicular structures are characteristic of the intermediate concentration regime. The data finally suggests that TEQ-bilayers are stable upon dilution, and that concentration has an effect on the vesicular size and morphology. Concerning the structures in Fig. 1, it should be noted that in contrast to earlier reports,[7,35-36,37] multi-lamellar vesicles comprising periodically stacked bilayers in the form of onions were not observed. Seth *et al.* also described a wide variety of bilayer assemblies (e.g. deflated unilamellar and stomatocytes) for crowded dispersions undergoing an electrostatic induced destabilization.[38] These transient structures were also observed, but in minute proportions.

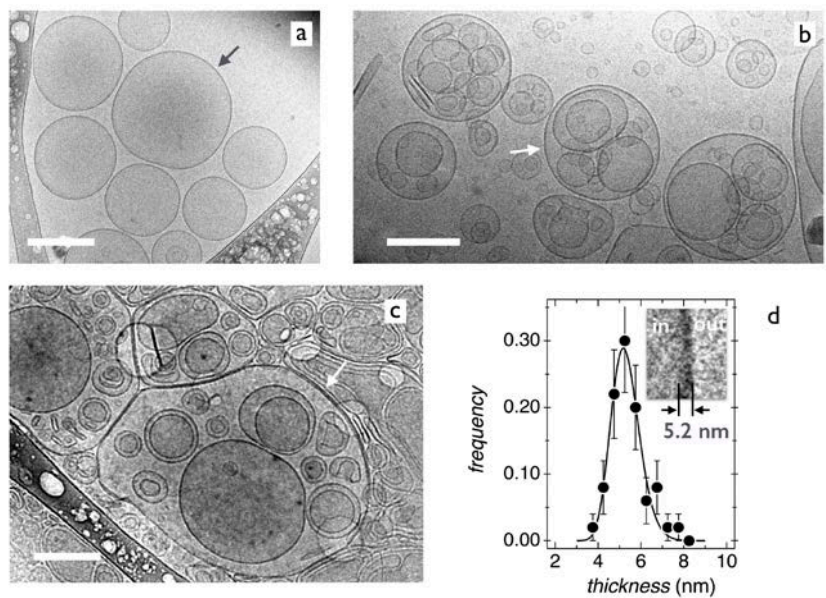

**Figure 1**: Representative cryogenic transmission electron microscopy (cryo-TEM) images of double-tailed cationic surfactant dispersions. Concentrations are 0.1 wt. % (a), 1 wt.% (b) and 4 wt. % (c). Dark and white arrows indicate the unilamellar and multivesicular vesicles respectively. The bars are 200 nm. d) Bilayer thickness distribution obtained from cryo-TEM image analysis. The average value and dispersity index are 5.2 nm and 0.12 respectively. Inset: Close-up view of the surfactant bilayer.

### III.1.2 - X-ray scattering

In Fig. 2a, the scattering intensity obtained at concentrations between 4 and 20 wt. % exhibits a $1/q^2$-decrease at low wave-vector that is typical for vesicular aggregates, followed by a pronounced oscillation around 1 nm$^{-1}$ and a smaller one at 3 nm$^{-1}$. These latter oscillations are characteristic of the surfactant bilayer form factor.[18,39-41] In this range, the intensity scales approximately with the concentration. Weak intensity scattering peaks are also observed in concentrated samples at 8, 12 and 15 wt. %, and they arise from structure factors of densely packed bilayers enclosed into multilamellar vesicles. In Supplementary Fig. S1, the scattering intensity obtained from TEQ dispersions at 8, 12, 15 and 20 wt. % was divided by the bilayer form factor (and normalized by the concentration), yielding the structure factor at the length scale of the bilayers. For the 8, 12 and 15 wt. % samples, three Braggs peaks at wave-vectors $q_i$ (i = 1 − 3) were observed, with $q_2/q_1$ = 2 ± 0.1 and $q_3/q_1$ = 3 ± 0.2. The derived interlamellar distance was found to decrease with concentration from 20.6, to 13.3 and 11.6 nm. Combined with the cryo-TEM data in the previous section, the x-ray results indicate that the microstructure changes with the concentration, from unilamellar vesicles in the dilute regime to multivesicular/multilamellar vesicles in the concentrated regime. It is probable that at the concentration at which TEQ softeners are formulated (12 wt. %), the three types of structures coexist.

Adjustment of the bilayer form factor using a two-level electronic density profile, one for the head groups and one for the alkyl chains of thickness $\delta_H$ and $\delta_T$ respectively[39,41] provides reasonable good fitting in the range 0.2 − 3 nm$^{-1}$. Thicknesses $\delta_H$ = 0.8 nm and $\delta_T$ = 1.5 nm were obtained for the two compartments, respectively, leading an overall layer thickness of $2(\delta_H + \delta_T)$ = 4.6 nm (Supplementary Information S2). These results are in agreement with literature data on similar surfactant[18] or phospholipids.[40] More elaborate models taking into account vesicle form factor and size dispersity would be required for adjusting the data over a more extended q-range.[18,40,42] Figs. 2b and 2c shows high wave-vector intensities obtained by WAXS at $T$ = 20 and 50 °C. Data at 2, 4 and 12 wt. % display pronounced peaks associated with an hexagonal order of the surfactant molecules within the layer.[18,42] These Braggs peaks indicate that the bilayers are in a gel phase at the two temperatures. The peak intensities decrease however from 20 to 50 °C, suggesting the nearing of the gel-to-fluid transition that was estimated independently from calorimetry at around 60 °C. Note that the gel phase ensures a strong stability of the vesicles in the conditions of use, both in concentration and in temperature.

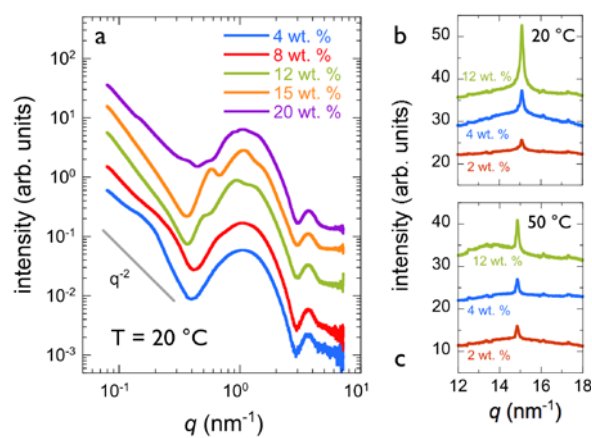

**Figure 2**: *Small-angle X-ray scattering from ethanaminium, 2-hydroxyN,N-bis(2-hydroxyethyl)-N-methyl-esters (TEQ) surfactant at concentrations 4, 8, 12, 15 and 20 wt. % (T = 20 °C). At low wave-vectors, the intensity decreases as $q^{-2}$, as expected for vesicles. The curves were shifted with respect to each other for clarity. b) Wide-angle X-ray scattering at T = 20 °C in the range 12 – 18 $nm^{-1}$ for TEQ 2, 4 and 12 wt. %. c) Same as in b) for T = 50 °C.*

## III.2 – Surfactant assembly at the micrometer scale

To study TEQ vesicles at larger length scale, phase-contrast optical microscopy experiments were performed. Figs. 3a displays a 20 × magnification image of the dispersion at the concentration at which it is formulated, c = 12 wt. %. It shows highly contrasted and densely packed vesicular objects with sizes up to 10 μm (arrows). At this concentration, the volume fraction is elevated[38,43] and most vesicle motions are frozen. This structural arrangement imparts to the fluid the rheological properties of a soft solid, characterized by a gel-like elastic complex modulus and a non-zero yield stress (Supplementary Information S3). The vesicular size distribution (Fig. 3b) is characterized by a high dispersity, and a significant number of objects are larger than 5 μm. In some cases, the large vesicles are not spherical and exhibit facets (Fig. 3a). This latter phenomenon is attributed to mechanical stresses generated during preparation. It is related to the liquid-crystalline order of the alky chains within the layer, as shown previously by wide-angle x-ray scattering.[18,42] Fig. 3c shows a microscopy image of a dilute TEQ solution at the concentration used in the washing machine (c = 0.024 wt. %). The dispersion was obtained by dilution of the 12 wt. % formulation with DI-water at pH 4.5. The figure reveals well-contrasted spherical particles of average size 1 μm (Fig. 3d) that are animated from rapid Brownian motions (see movie#1 and #2 in Supplementary Information). This finding confirms that TEQ vesicles are stable upon dilution. Dilute surfactant dispersions were further investigated by dynamic light scattering. Fig. 3e and 3f display the second-order correlation functions $g^{(2)}(t)$ as a function of delay time and the corresponding hydrodynamic diameters *versus* concentration, respectively. The autocorrelation functions display a unique relaxation mode at all concentrations. In Fig. 3f, $D_H$ is found to increase with concentration, from 400 to 750 nm. These findings suggest that as in cryo-TEM and x-ray scattering, dilution tends to reduce the average vesicular size. This size reduction observed is due to counterion mediated osmotic effects induced by dilution.[9,44] Electrophoretic mobility measurements on dilute dispersions resulted in zeta potential $\zeta$ = +65 mV, indicating that the vesicles are positively charged. In conclusion to this part, we have found that TEQ softener surfactants assemble into vesicles, and that the bilayer structure remains stable upon temperature (T < 50 °C) and concentration changes. The local membrane organization depends however on the concentration, as structures ranging from unilamellar to multivesicular and multilamellar vesicles were disclosed.

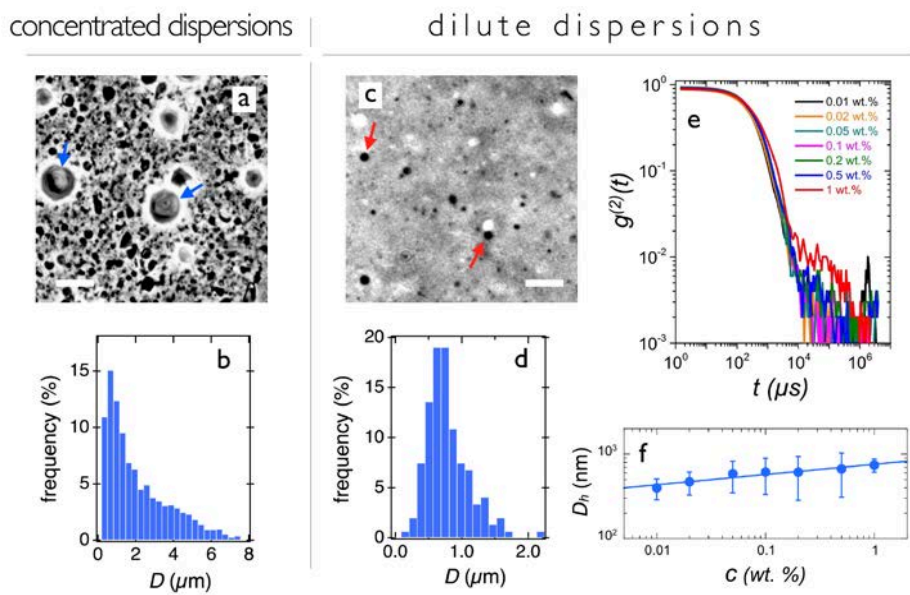

concentrated dispersions | dilute dispersions

**Figure 3**: *TEQ surfactant dispersions studied by phase-contrast microscopy and dynamic light scattering. a) Microscopy image of a 12 wt. % dispersion. b) Vesicular size distribution corresponding to the data in Fig. 3a. c) Microscopy image of a 0.024 wt. % dispersion. d) Vesicle size distribution corresponding to the data in Fig. 3c. e) Second-order auto-correlation functions of dispersions between 0.01 and 1 wt. %. f) Hydrodynamic diameter as a function of the concentration corresponding to the data in Fig. 3e. Bars in Fig. 3a and 3c are 10 μm.*

## III.3 – Cellulose nanocrystals

Cellulose nanocrystals studied in this work were obtained from cotton microfibril oxidation and subsequent mechanical treatment.[2,29-30] Fig. 4a shows a cryo-TEM image of the anisotropic objects dispersed in water at pH 4.5. Longitudinal and transverse dimensions are 17 ± 4 nm and 180 ± 30 nm. The cellulose nanocrystals are similar in shape and size to those reported in the literature.[22,29] To study the nanocrystalline cellulose/surfactant interactions, the state and stability of the dispersions over time were first evaluated. Using light scattering, the CNC dispersion prepared at pH 4.5 and at a concentration of 0.1 wt.% displays a single relaxation mode in the autocorrelation function. $g^{(2)}(t)$ decreases rapidly above delay times around $10^3$ μs, indicating that the nanofibers are not aggregated. Light scattering was carried out using non-polarized light for the incident and scattered beams, and the mode in Fig. 4b is a combination of the orientational and translational diffusion.[45-46] A direct analysis yields an intensity distribution centered around 120 nm, in agreement with the cryo-TEM data. Electrophoretic mobility experiments resulted in zeta potential $\zeta$ = -38 mV, indicating that the nanocrystals (as cotton) are negatively charged in the conditions of use.

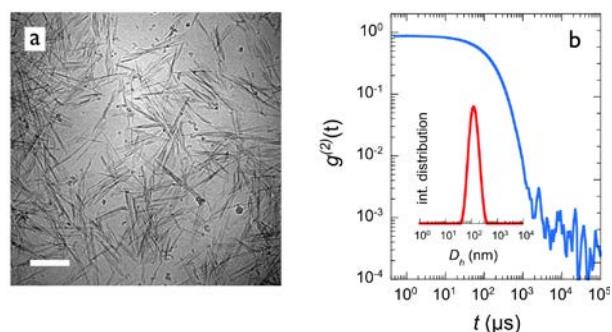

**Figure 4**: *a) Cryogenic transmission electron microscopy (cryo-TEM) image of nanocrystalline cellulose. The bar is 100 nm. b) Second-order autocorrelation function of the scattered light showing a single relaxation mode associated to fiber diffusion. Inset: related intensity distribution.*

## III.4 – Cellulose nanocrystal/surfactant interaction

### III.4.1 – Light scattering and zetametry

Nanocrystalline cellulose/surfactant dispersions were formulated by mixing stock solutions at different volumetric ratios between $10^{-2}$ and $10^2$. Figs. 5a and 5b show the Rayleigh ratio and hydrodynamic diameter for dispersions prepared at 0.01 wt. % total concentration. Both data exhibit a marked maximum at $X_{Max}$ = 0.3. The continuous green line in Fig. 5a is calculated assuming that $\mathcal{R}(X)$ is the sum of the Rayleigh ratios $\mathcal{R}_{CNC}$ and $\mathcal{R}_{TEP}$ weighted by their respective concentrations, where $\mathcal{R}_{CNC}$ and $\mathcal{R}_{TEP}$ are the Rayleigh ratios of the nanocrystalline cellulose and surfactant dispersions.[32-33,47] In Figs. 5a, $\mathcal{R}(X)$ is found to be higher than the predictions for non-interacting species. The observation of a maximum indicates that CNC/surfactant aggregates are formed and that the aggregates are characterized by a fixed stoichiometry.[33,48-49] Comparing the present data with earlier reports on inorganic particles and synthetic polymers,[28,33,47-48,50-52] the scattered intensity found here is high ($\mathcal{R} > 10^{-3}$ cm$^{-1}$) and suggests a strong interaction between the two species. The hydrodynamic diameter data in Fig. 5b are in agreement with this conclusion, as $D_H$ reaches maximum values of several microns.

Fig. 5c shows the distribution intensities *versus* zeta potential $\zeta$ obtained from electrophoretic mobility measurements between $X$ = 0 and $X$ = 100. The intensities exhibit well-defined peaks that shift progressively from negative ($\zeta$ = - 38 mV) for pure cellulose nanocrystals to positive ($\zeta$ = + 65 mV) for pure TEQ solutions. The point of zero charge is found at $X$ ~ 1 (Fig. 5d), *i.e.* slightly higher than the peak position found in light scattering. These results are interpreted in terms of electrostatic complexation, associated with the process of charge titration and neutralization.[48-49] In this scenario, the aggregates formed at equivalence have a zero surface charge and grow rapidly in size. Apart from the stoichiometric, the aggregates are charged (negative for $X$ < 1 and positive for $X$ > 1) and repel each other, preventing their further growth and sedimentation. Fig. 5e shows images of cuvettes containing mixed dispersions prepared between $X$ = 0.05 and 5 and at total concentration $c$ = 0.1 wt.% (*i.e.* 10 times that of light scattering). There, the CNC/surfactant aggregates appear as a turbid phase that tends to sediment over time. The figure confirms the light scattering data, namely that the more turbid samples are around $X$ = 0.3. The features in Fig. 5 were observed with other colloidal systems, including synthetic and biological polymers, phospholipid vesicles and surfactants and they are now considered as typical for electrostatic based interaction.[32-33,48,50-53]

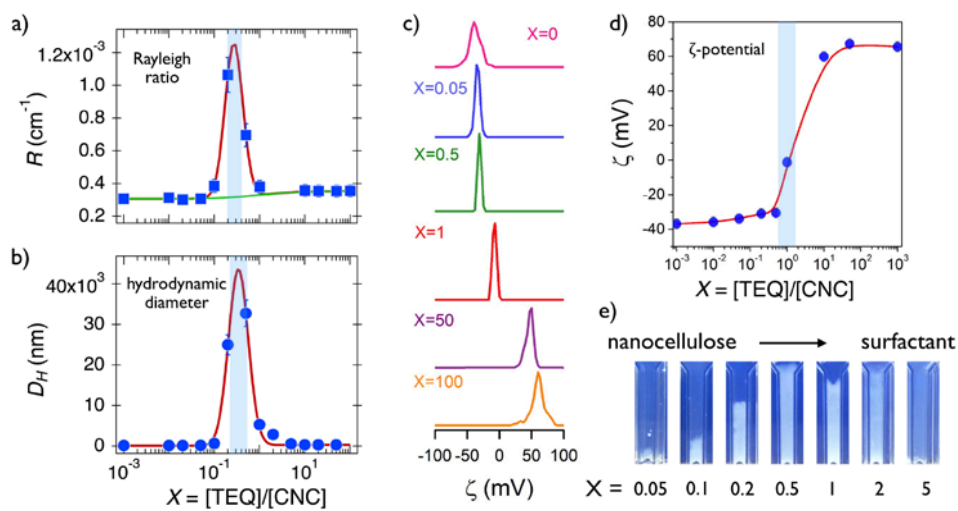

**Figure 5:** *a) Rayleigh ratio $\mathcal{R}(X)$ and b) hydrodynamic diameter $D_H(X)$ for mixed cellulose nanocrystal/surfactant dispersions as a function of the mixing ratio $X$. Temperature is 25 °C and concentration $c$ = 0.01 wt. %. In a), the continuous line in green is calculated for non-interacting species. The continuous lines in red are guides to the eyes. c) Intensity distributions as a function of the zeta potential $\zeta$ for samples between $X$ = 0 and $X$ = 100 showing a displacement of the peak from negative (pure nanocrystalline cellulose) to positive (pure TEQ vesicles) zeta values. d) Zeta potential results $\zeta(X)$ corresponding to the samples in c). The continuous line in red is a guide to the eyes. e) Images of mixed dispersions at $c$ = 0.1 wt. %.*

To validate the Continuous Variation Method and provide a negative control experiment, the complexation properties between nanocrystalline cellulose and an uncharged guar hydrocolloid was investigated. The guar put under scrutiny is a high molecular weight gum used as viscosifier in hydraulic fracturing applications. It is characterized by a hydrodynamic diameter of 200 nm and a zeta potential $\zeta$ = - 3 mV, indicating that the hydrocolloid is not charged. Supplementary Fig. S4 displays the Rayleigh ratio $\mathcal{R}(X)$ and hydrodynamic diameter $D_H(X)$ for CNC/guar dispersions *versus* mixing ratio. The data vary smoothly from one stock dispersion to the other, showing no extra scattering peak. Moreover, the $\mathcal{R}(X)$-variation is well accounted for by the non-interacting species model. It is concluded that the anionic nanocellulose crystals and uncharged guar hydrocolloid do not mutually interact.

### III.4.2 – Fluorescence microscopy

To further analyze the structure of the mixed aggregates and examine the hypothesis of vesicles remaining intact upon mixing,[7] the TEQ bilayers were labeled with the fluorescent dye PKH67. PKH67 is a tag molecule used in cellular biology and known to insert into phospholipid membranes (inset in Fig. 6a). In dilute surfactant solutions, the vesicles observed in 60× phase-contrast microscopy are identified as separated objects at the glass slide interface (Fig. 6a). With fluorescence, the vesicles appear as bright spots co-localized with those observed in phase-contrast, as shown in Fig. 6b. An analysis of the merge image (Fig. 6c) leads to the result that the majority of treated vesicles (> 80%) have been successfully labeled with the dye. Fig. 6d displays a vesicular size distribution centered at 760 nm.

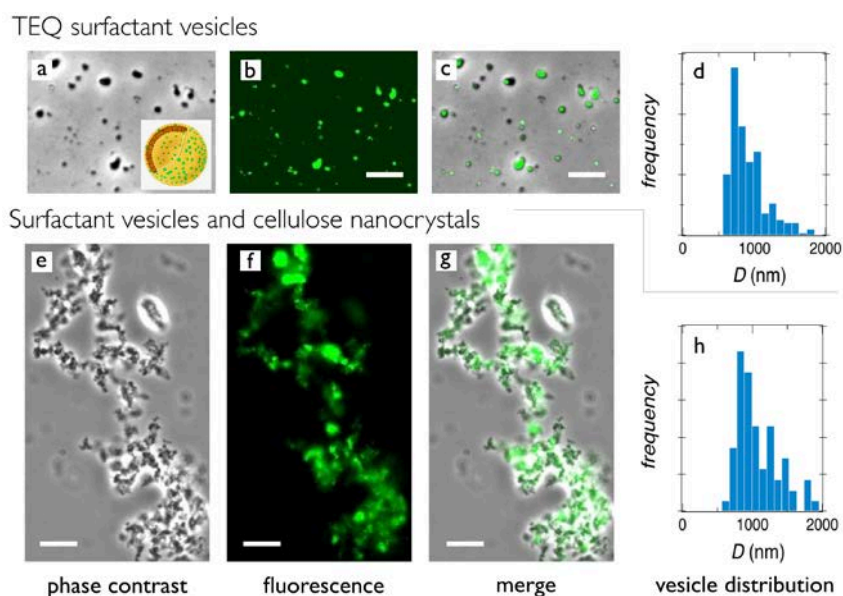

**Figure 6:** *Phase-contrast and fluorescence optical microscopy showing surfactant vesicles (a-c) and mixed cellulose nanocrystal/vesicle aggregates (e-g). Samples were prepared at c = 0.05 wt. % and X = 0.2. Images in c) and g) display the merges of the different structures found in phase-contrast and fluorescence. The vesicle size distributions shown in d) and h) are centered at 760 and 900 nm, respectively (with a dispersity of 0.26). The bars are 5 µm.*

When complexed with nanocrystalline cellulose, large aggregates form as a result of the strong electrostatic interaction. Fig. 6 displays an extended field of view comprising a large CNC/vesicle aggregate deposited on the glass slide. The experimental conditions for this image are a concentration of 0.05 wt. % and a mixing ratio X = 0.2, *i.e.* at the onset of the light scattering peak. In phase-contrast (Figs. 6e) and in fluorescence (Fig. 6f), the images show a large number of assembled vesicles sticking to each other, with no detectable diffusion. The merged image in Fig. 6g confirms the good co-localization of the vesicles seen with the two techniques. Fig. 6h illustrates the vesicle size distribution retrieved from the data in Fig. 6e. It exhibits a maximum at 900 nm in good agreement with that of single uncomplexed vesicles. These results suggest that TEQ vesicles preserve their integrity after interacting with the cellulose nanocrystals. To confirm these findings, cryo-TEM was attempted on the same mixed dispersion (Supplementary Information S5). However mixed CNC/vesicle aggregates could not be observed so easily with this technique. It was concluded that the micron-sized aggregates discussed previously were removed when blotting the dispersion droplet from the grid. This would explain the low number of aggregates found, and the difficulty to obtain conclusive data.

## IV - Conclusion

Here we apply a facile and accurate technique to evaluate the interaction of surfactant conditioners with cotton fibers in the context of household and personal care applications. The technique borrows its principle from the Continuous Variation Method developed almost a century ago and adapted for polymer and colloid solutions.[24-26] To mimic the surfactant absorption at the cotton surface, we use cellulose nanocrystal dispersions and investigate bulk solutions. The surfactant studied is a double-tailed cationic surfactant present in conditioners, which is known to exhibit good deposition performances on fabrics. At first, the individual components, surfactant and nanocrystalline cellulose are characterized. Cryo-TEM experiments show that in the conditions of use the cationic surfactant self-assembles into unilamellar, multivesicular and multilamellar vesicles according to the concentration. In the conditions tested *i.e.* between 0.01 to 20 wt. %, the vesicle sizes are distributed and vary between 100 nm and 10 μm. The cationic vesicles are also found to be stable upon temperature changes (T < 50 °C). Wide-angle x-ray scattering reveals the presence of Braggs peaks associated with an hexagonal order of the surfactant molecules within the bilayer.

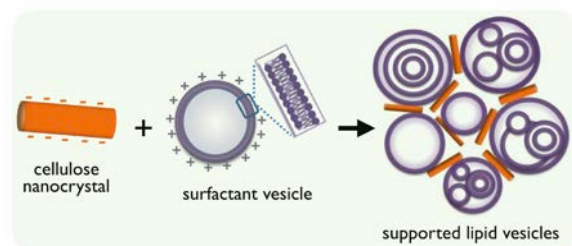

**Figure 7:** *Schematic representation of the electrostatic co-assembly of surfactant vesicles and cellulose nanocrystals.*

As for cellulose nanocrystals, cryo-TEM provides a direct visualization of the rod-shaped particles and confirms their lath morphology. As cotton, the laths are negative charged and display a zeta potential of -38 mV. Mixed CNC/surfactant dispersions exhibit a unique feature as a function of the mole fraction ratio. At the point of zero charge, the dispersions exhibit a strong scattering peak indicative of the formation of micron-sized aggregates. Such findings have been found in a wide variety of colloidal systems and represent the signature of electrostatic-based interactions.[33,48,50-52,54] Fluorescence microscopy supports the conclusion that the vesicles adsorbed at the cellulose nanocrystal interface remain intact, a result that could be ascribed to the bilayer stability. A schematic representation of the nanocrystal/vesicle co-assembly is provided in Fig. 7. In conclusion, the technique developed here could be exploited to rapidly assess the deposition efficiency of fabric conditioners on cotton by changing the amount and nature of chemicals in the formulations. For an efficient screening, it could be also automated using fluidics, multi-sample support and high-speed reading capacity.

## Supporting Information

The Supporting Information includes sections on the determination of the bilayer structure factor (S1) and bilayer thickness (S2) from x-ray scattering and on the linear and nonlinear rheology characterization of the surfactant concentrated dispersions (S3). A control experiment with guar hydrocolloids using the Continuous Variation Method is shown in S4, whereas S5 displays some cryo-TEM images of cellulose nanocrystal/vesicle aggregates.

# THE JOURNAL OF PHYSICAL CHEMISTRY B

# Acknowledgements

The research performed in this work was funded by Solvay. ANR (Agence Nationale de la Recherche) and CGI (Commissariat à l'Investissement d'Avenir) are gratefully acknowledged for their financial support of this work through Labex SEAM (Science and Engineering for Advanced Materials and devices) ANR 11 LABX 086, ANR 11 IDEX 05 02.

## TOC Graphic

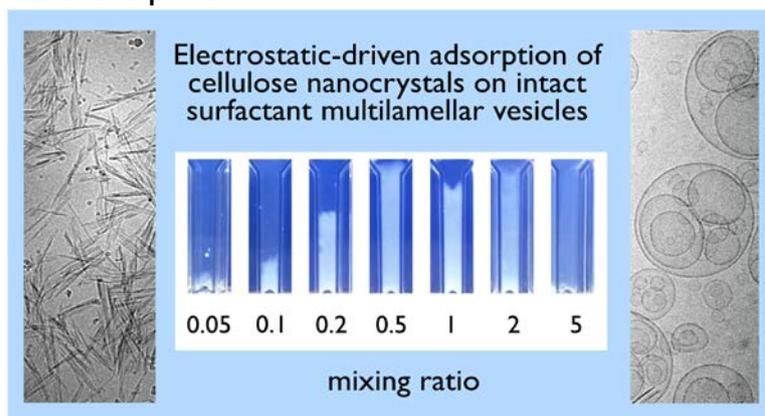